\documentclass[iop]{emulateapj}

\usepackage{graphicx}

\shorttitle{Herschel Imaging of the HR 8799 Debris Disk}
\shortauthors{Matthews et al.}

\begin{document}

\title{Resolved Imaging of the HR 8799 Debris Disk with Herschel\footnote{{\it Herschel} is an ESA space observatory
with science instruments provided by European-led Principal
Investigator consortia and with important participation from NASA.} }

\author{Brenda C. Matthews\altaffilmark{1,2}, Grant Kennedy\altaffilmark{3}, Bruce Sibthorpe\altaffilmark{4}, Mark Booth\altaffilmark{5,2,1},Mark Wyatt\altaffilmark{3}, \\ Hannah Broekhoven-Fiene\altaffilmark{2,1}, Bruce Macintosh\altaffilmark{6,7}, Christian Marois\altaffilmark{1,2}}

\altaffiltext{1}{National Research Council of Canada Herzberg Astronomy \& Astrophsyics, 5071 W.\ Saanich Road, Victoria, BC, V9E 2E7, Canada}
\altaffiltext{2}{Department of Physics \& Astronomy, University of Victoria, 3800 Finnerty Rd, Victoria, BC, V8P 5C2, Canada}
\altaffiltext{3}{Institute of Astronomy, University of Cambridge, Madingley Road, Cambridge, United Kingdom, CB3 0HA}
\altaffiltext{4}{SRON Netherlands Institute for Space Research, PO Box 800, NL-9700 AV Groningen, the Netherlands}
\altaffiltext{5}{Instituto de Astrof\'isica, Pontificia Universidad Cat\'olica de Chile,  Vicu\~na Mackenna 4860, 7820436 Macul, Santiago, Chile}
\altaffiltext{6}{Lawrence Livermore National Labs, 7000 East Ave, Livermore, CA 94550, U.S.A.}
\altaffiltext{7}{Department of Physics \& Kavli Institute for Particle Astrophysics and Cosmology, Stanford University, Palo Alto, CA U.S.A.}

\begin{abstract}
We present {\it Herschel} far-infrared and submillimeter maps of the debris disk associated with the HR 8799 planetary system.  We
resolve the outer disk emission at 70, 100, 160 and 250 \micron\ and
detect the disk at 350 and 500 \micron.  A smooth model explains the observed disk emission well.  We observe no obvious clumps or asymmetries associated with the trapping of planetesimals that is a potential consequence
of planetary migration in the system.  We estimate that the disk eccentricity must be $<0.1$.  As in previous work by \cite{su09}, we find a disk with three components: a warm
inner component and two outer components, a planetesimal belt extending from 100 - 310 AU, with
some flexibility ($\pm 10$ AU) on the inner edge, and the external
halo which extends to $\sim 2000$ AU.  We measure the disk inclination
to be $26 \pm 3^\circ$ from face-on at a position angle of 64$^\circ$
E of N, establishing that the disk is coplanar with the star and
planets.  The SED of the disk is well fit by
blackbody grains whose semi-major axes lie within the planetesimal
belt, suggesting an absence of small
grains. The wavelength at which the spectrum steepens from blackbody,
$47 \pm 30$
\micron, however, is short compared to other A star debris disks,
suggesting that there are atypically small grains likely populating the halo. The PACS longer wavelength data yield a
lower disk color temperature than do MIPS data (24 and 70 \micron), implying two distinct halo dust grain populations.
\end{abstract}

\keywords{(stars:) circumstellar matter, planetary systems, individual (HR 8799)}

\section{Introduction}

True Solar system analogues, systems of multiple planets with evidence
for warm and cold dust disk components, have few representatives
outside our own. A recently recognized exception is the HR 8799 (an
A5V star at 39.9 pc) system which hosts a directly imaged multi-planet
system and shows evidence for warm and cold dust disk components.
Four planetary companions have been detected around HR 8799 at
projected separations of 15, 24, 38 and 68 AU
\citep{mar08,mar10}.  Evidence for a circumstellar dust disk has been
known for some time \citep{sad86}, and {\it Spitzer} observations have
revealed that the disk must contain multiple dust components
\citep{su09,rei09}.  \cite{su09} found evidence for three disk
components: a warm central component, an outer ``cold component''
extending from 90 - 300 AU, and an external halo of small grains
extending to an outer radius of 1000 AU.  The relatively young age of
this system \citep[the most recent estimate gives an age of 20-50
Myr,][]{mar10} makes it especially intriguing, since this is the
currently favored epoch for the formation of terrestrial planets
\citep{cha01,ray04,mel10}.

\begin{deluxetable}{cccc}
\tabletypesize{\scriptsize}
\tablecaption{Observations Log}
\tablewidth{0pt}
\tablehead{
\colhead{Obs. ID} & \colhead{Observing date} & \colhead{Mode} & \colhead{Duration} }
\startdata
1342223848/9  & 9/10 July 2011 & PacsPhoto 70/160 & 9018 $\times$ 2 \\
1342223850/1 & 10 July 2011 & PacsPhoto 100/160 & 7890 $\times$ 2 \\
1342234924-28 & 18 Dec 2011 & SpirePhoto & 169 $\times$ 5 
\enddata
\label{obslog}
\end{deluxetable}

\begin{deluxetable*}{ccccc}
\tabletypesize{\scriptsize}
\tablecaption{Observations Summary}
\tablewidth{0pt}
\tablehead{
\colhead{Wavelength} & \colhead{FWHM} & \colhead{Integration time} & \colhead{instrumental noise} & \colhead{1-$\sigma$ rms noise level} \\
\colhead{[\micron]} & \colhead{[\arcsec]} & \colhead{[hours]} & \colhead{[mJy beam$^{-1}$]} & \colhead{[mJy beam$^{-1}$]} }
\startdata
70  & 5.5 x 5.8  & 5.1 & 0.4 & 0.338 \\
100 & 6.7 x 6.9 & 4.45 & 0.5 & 1.43 \\
160 & 10.7 x 12.1 & 9.55 & 0.6 & 8.90 \\
250 & 18.7 x 17.5 & 0.3 & 4.0 & 25.16 \\
350 & 25.6 x 24.2 & 0.3 & 3.4 & 29.01 \\
500 & 38.2 x 34.6 & 0.3 & 4.8 & 20.70 
\enddata
\label{obsdetails}
\end{deluxetable*}

We also have an ideal viewing angle toward this system, as the orbital
plane of the planets and disk appear to lie very close to the plane of
the sky \citep{mar08,laf09,rei09,su09,fab10,sou11}, although the inclination of the star has been estimated to be
in excess of 40\degr\ by \cite{wri11}.  A face-on geometry
makes it possible to observe azimuthal variations in the disk
structure, as well as radial ones (e.g., $\epsilon$ Eri,
\cite{gre98,gre05} and Vega, \cite{hol98}, though see \cite{bac09} and \cite{hug12}).  The presence of
so many giant planets in a system makes it inevitable that the debris
disk has been sculpted by gravitational perturbations.
\cite{wya99} and \cite{mou97} show that the planets' secular
perturbations could impose eccentricity or inclination on the disk
\citep[e.g., $\beta$ Pic, ][]{lag12}, or migration of the planets
could have trapped material into resonance \citep{wya03}.

The combination of four massive, coeval, and spectroscopically
characterizable planets \citep{kon13,bow10}, together with the dust disk, makes this
system a ``Rosetta Stone'' for planet formation. The three outer
planets have significant gravitational interactions
\citep{fab10,rei09,goz09,goz13}; the age of the system requires it to be stable over
$10^8$ year timescales. This in turn requires either a very finely
tuned orbit for the nominal 5-7 $M_J$ planets, or low planet masses.
Precision astrometry \citep{kon09,kon10} being carried out at the Keck
observatory can constrain these orbital scenarios even over partial
orbits. However, the outermost planet, ``b'', with its long period, is
the most weakly constrained. Future resolution of the inner edge of
the planetesimal belt will provide significant constraints on
outer-planet inclination given the coplanarity of the planets and disk
discussed in this work. Resolving the inner edge will also provide constraint on the eccentricity of the disk to a simultaneous astrometric and
dynamical fit. Finally, such high resolution images can provide mass information in the same way that the dust
rings have been used around $\epsilon$
Eri \citep{qui02} and around Fomalhaut to infer an orbit and mass of
Fomalhaut b, though it now appears to apply to a hypothetical
Fomalhaut c \citep{chi09,kal13}.

We present higher resolution far-infrared data than was obtainable by
{\it Spitzer} toward HR 8799 from the {\it Herschel Space
Observatory's} PACS \citep[Photodetector Array Camera and
Spectrometer,][]{pog10} instrument at 70, 100 and 160 \micron. In
addition, we present lower resolution submillimeter images from the
SPIRE \citep[Spectral and Photometric Imaging REceiver][]{gri10}
camera at 250, 350 and 500 \micron.  These data were taken during the
{\it Herschel} Open Time Cycle 1 and reveal a well-resolved disk around HR
8799 in the far-infrared.  We describe our observations in $\S$
\ref{obs}, describe the data reduction in $\S$ \ref{datared} and
present the data in $\S$ \ref{results}. In $\S$ \ref{ss:models}, we present models of the resolved images. 
In $\S$ \ref{discussion}, we explore the implications of these data. We summarize our findings in 
$\S$ \ref{conc}.

\section{Observations}
\label{obs}

The record of the observing log is shown in Table \ref{obslog}.

The PACS maps were made using the mini-scan mapping mode \citep{pog10}
at a scan speed of 20\arcsec/s across the target in 10 scan legs,
separated by a scan line cross-step of 4\arcsec. Each scan leg is
3\arcmin\ long. At 70 \micron, there are 64 repeats of each map, with
the observations paired into two scanning angles: 70 and 110 degrees
(with respect to the instrument reference frame). At 100 \micron, 56
maps are required, with 28 at each of the two scanning angles
specified above. The total integration time on the target is 5.1 hours
with PACS 70/160 and 4.45 hours with PACS 100/160. The 160 \micron\
data were obtained at the same time as the 70 \micron\ and 100
\micron\ data. The combined 160
\micron\ map from both observations just reaches the 160 \micron\
confusion limit.

SPIRE observes all three submillimeter bands simultaneously to
the confusion limit in 0.3 hour of time. Five maps were made in
small-map mode with a dither set between each map of 12\arcsec\ (twice
the nominal 250 \micron\ pixel scale). The concatenation of these
observations provides a more uniform coverage of the disk emission than
a single longer integration and allows us to use a finer pixel scale in the 
output maps. 

The total mapped area of the PACS observations is 3.5\arcmin\ $\times$
6.5\arcmin, but the area with high coverage, i.e., the ``map center'',
corresponds to all regions that have at least 80\% of the peak
coverage, and high integration time is restricted to 1\arcmin\
$\times$ 2\arcmin.  The details of the observations and the resulting
rms noise estimates at each wavelength are summarized in Table
\ref{obsdetails}.  The high S/N coverage area of the SPIRE maps is
4\arcmin\ $\times$ 4\arcmin.

Noise measurements are taken from within a 2\arcmin\ $\times$
2\arcmin\ area centred on the source.  The noise is estimated by
taking the RMS of a series of many summed apertures across the field,
with each aperture having an area equal to that of a
diffraction-limited Gaussian beam.  Note that the high noise values at
and longward of 160 \micron\ are a result of the high cirrus confusion
in the region.  The fact that the noise peaks at around 250-350
\micron, i.e. the peak of the ISM dust emission, and then drops at 500
\micron\ is indicative that this is the primary source of noise at these
wavelengths.

\section{Data reduction}
\label{datared}

The data were reduced using version 7 of the {\it Herschel} interactive
pipeline environment \citep[HIPE][]{ott10} software, implementing 
versions 32
and 8.1 of the PACS and SPIRE calibration products respectively.  The
standard processing steps were followed, with an additional jack-knife
deglitching step implemented in the PACS data reduction, removing any
artefacts with a significance greater than 5-$\sigma$ in the
jack-knifed image.  This step was performed as an additional
deglitching precaution only.

The PACS time-line data were high-pass filtered to reduce the impact of 
1/$f$ noise (see Section~\ref{filtering} for details), and converted to a
map using the HIPE `photProject' task. Maps were likewise made from
the SPIRE data using the `naiveMapper' task.  No filtering was
performed on the SPIRE data.

The maps were created with pixel scales a fraction of the beam size.
The sizes used are 1\arcsec\ for all PACS maps, 3\arcsec\ for 250 \micron, 4\arcsec\ for 350 \micron, and
5\arcsec\ for 500 \micron.  The SPIRE bands are highly
oversampled due to the dithering used during observations, which
allows us to produce maps with a relatively fine pixel scale.

\subsection{Filtering effects}\label{filtering}

During processing the PACS time-ordered data were high-pass
filtered to mitigate low 1/$f$ frequency noise.  Filtering was
performed on each bolometer individually, passing all scales above a
user-given input temporal scale.  Initially blind filtering was
performed and a map made from these filtered data.  This map was used
to identify all structure with a significance greater than or equal to
3-$\sigma$, from which a filter mask is created.  The original data
were then filtered again, this time masking regions containing real
structure, and the final map was projected.  Masking of map sources is
required to remove ringing style filter artefacts along the telescope
scanning directions.

Even though sources in the map were masked during filtering, the
expected extent and interest in the low-level emission from the
diffuse outer disk meant that it was critical to quantify the
impact of the filtering on the measured disk structure.  For example,
since the filtering is performed in the scan direction only, a small
filter scale could truncate the disk along the scan direction, making
a face-on disk appear to be inclined.  Even though two scanning
directions were used to perform these observations their separation was
only $40 ^\circ$, making disk truncation a possibility.  A large
filter scale improves source fidelity, but it also passes equivalently
larger angular scales to the map, resulting in an increased noise
level.

We began by producing reference maps with a filter scale of
66\arcsec\ (16 frames), and generated additional maps using
increasingly large filter scales.  The integrated flux density on the
source was then measured using a circular aperture with a radius of
30\arcsec\ in each map.  The background was subtracted using an annulus
centred on the source position, with inner and outer radii of 45\arcsec\ and
60\arcsec\ respectively.  The measured flux density from each map was
normalised to that measured in the reference map.  The rate of change
of normalised flux density with filter scale was initially high, and
decreased as the filter scale became large relative to the size of the
disk.  Following this analysis we chose to use maps created with a
filter scale of 402\arcsec\ (100 frames) where the rate of change was
small for all bands.

While this larger filter scale results in an increased noise level,
the increase in source flux is similar, meaning that there is little
loss in signal-to-noise ratio.  In addition, the improvement in source
structure fidelity is significant, and necessary for accurate disk 
modelling.

\section{Results}
\label{results}

\begin{figure*}
\vspace*{14cm}
\includegraphics{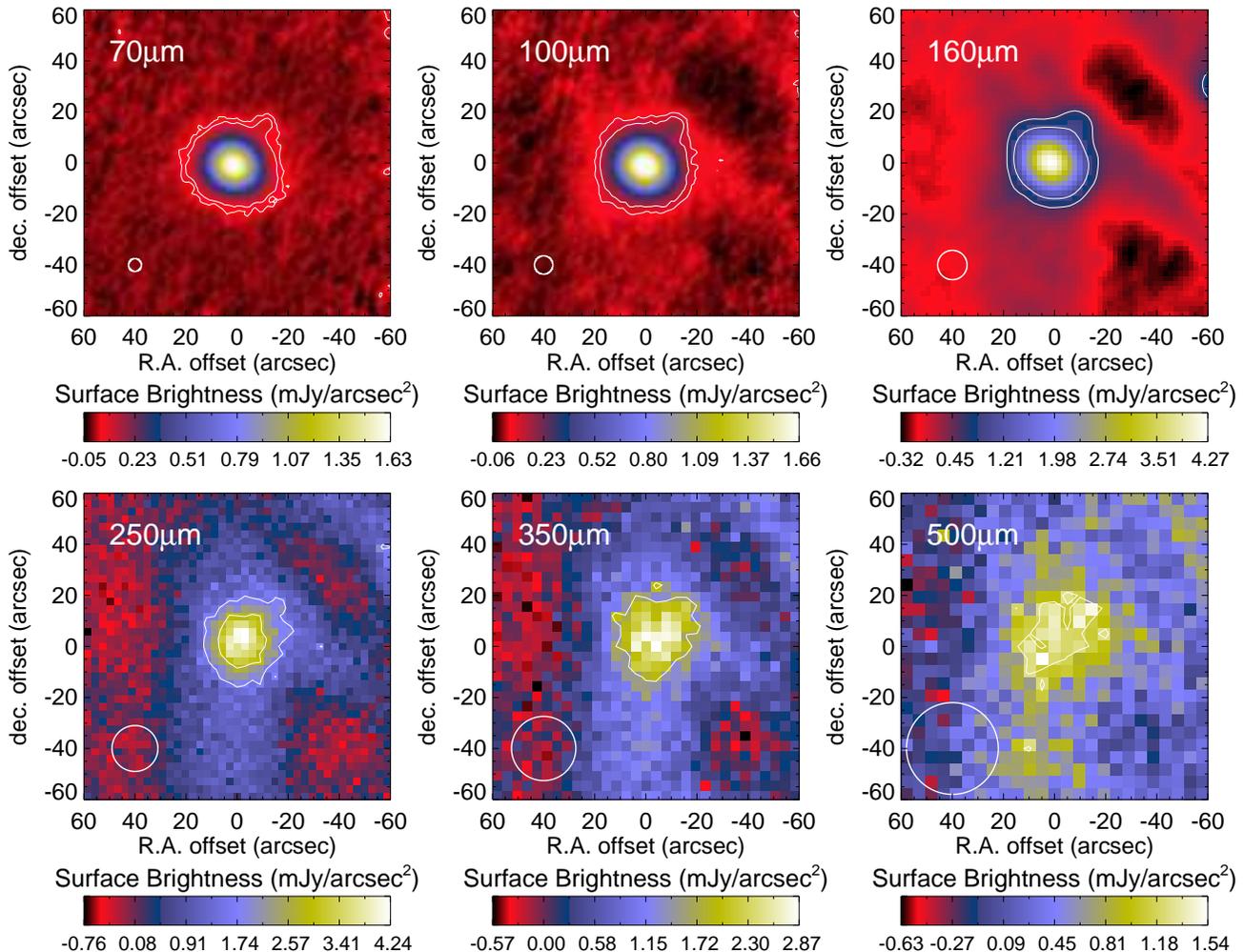}
\caption{Far-infrared and submillimeter maps of HR 8799 from PACS and SPIRE. 
Different surface brightness scales are used for each map, and the
pixel scales used are 1\arcsec\ pixels at 70 and 100 \micron,
2\arcsec\ pixels at 160 \micron, and 3, 4 and 5\arcsec\ pixels at 250,
350 and 500 \micron, respectively.  These pixels are small for SPIRE
data but the dithering used during these observations makes such fine
sampling possible.}
\label{images}
\end{figure*}

\begin{figure}
\vspace*{8cm}
\includegraphics{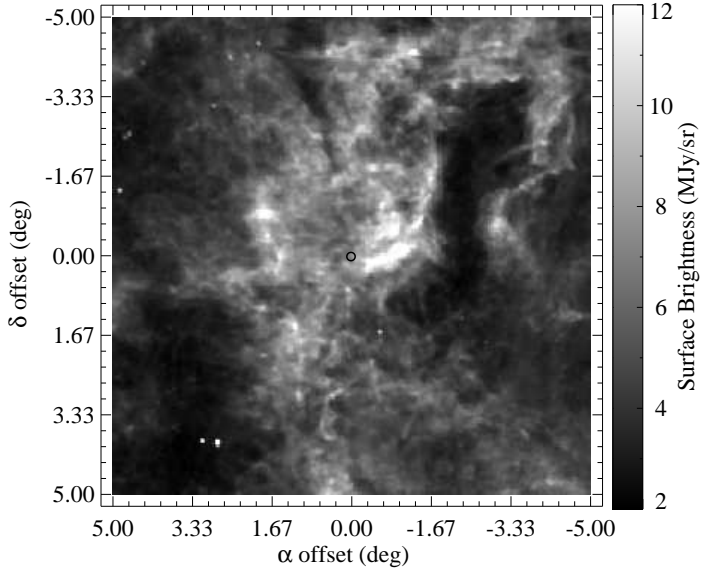}
\caption{IRAS 100 \micron\ image of the region surrounding HR 8799, 
marked at the field center by a circle.}
\label{irasimage}
\end{figure}

\begin{figure}
\vspace*{9cm}
\includegraphics{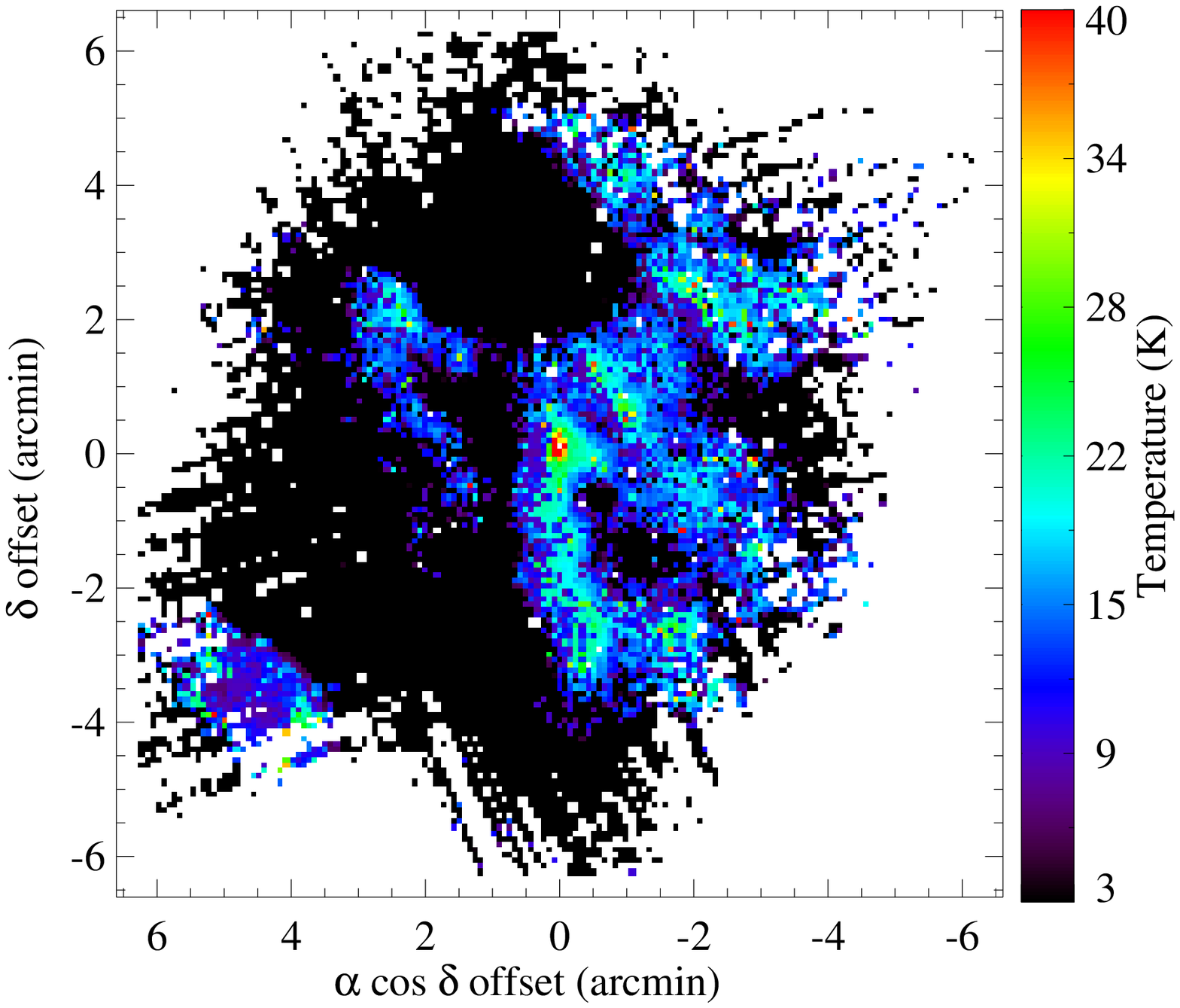}
\caption{Dust temperature map derived from the SPIRE data, illustrating the cold large-scale emission from a background molecular cloud along the line of sight to the position of HR 8799. }
\label{SPIRE3col}
\end{figure}

Figure \ref{images} shows the measured emission on the sky toward HR
8799.  HR 8799 is detected at all wavelengths.  A compact background
source, also detected at 24 and 70 \micron\ by \cite{su09}, is
detected $\sim 15$\arcsec\ northwest of the disk emission and is
resolved from HR 8799 at all PACS wavelengths.  
The images from 160 -- 500 \micron\ show asymmetric, large-scale
emission in addition to the HR 8799 disk emission. Figure
\ref{irasimage} shows an IRAS 100 \micron\ map of the sky around HR
8799. The significant large scale background emission underlying the
HR 8799 position arises from a (most likely background given the
proximity of HR 8799 at 39 pc) dust cloud.  We have used our SPIRE
data to make a temperature map of the large-scale emission by fitting
SEDs to each pixel in the SPIRE maps (convolved to common resolution),
shown in Figure \ref{SPIRE3col}.  The position of HR 8799 shows up as
a small region of warmer dust amid the $\lesssim 20$ K dust of the
background cloud.

The emission from the HR 8799 disk is clearly resolved at all PACS
wavelengths, but the impact of the underlying cirrus emission
increasingly affects interpretation of the data at and longward of 160
\micron.

\subsection{Photometry}\label{ss:phot}

Emission from the HR 8799 system was measured using aperture
photometry. These apertures are necessarily large to include as much
disk emission as possible, so at 70, 100 and 160 $\mu$m also include
emission from the NE background source. We used apertures of 30, 30
and 26\arcsec\ for 70, 100 and 160 $\mu$m, repectively. The aperture
at 160 $\mu$m was chosen to minimize the contribution from the
large-scale background cloud, but is large enough to encompass the point-like background source. At
this wavelength the high overall background level means that the flux density measured
depends strongly on the aperture size, an issue we revisit
below. Flux densities for the background source were estimated during the
modelling process using PSF fitting and were subsequently subtracted
from the aperture measurements.

An additional issue arises due to the high-pass filtering, which
removes emission on large spatial scales. While we have shown that more
disk emission is not gained by increasing the filter scale, it is
likely that emission is still missing from the aperture photometry
because the PACS beam has significant emission on large spatial
scales.  For example, the PACS documentation shows that at 100 $\mu$m
with a filter scale of 100\arcsec, about 10\% of the flux from a point
source lies beyond 30\arcsec. This emission cannot be detected with typical
observations\footnote{The extended emission was characterised by
observing Mars, which was heavily saturated near the peak but allowed
radial profiles of the outer parts of the beam profile to be derived.}
because the extended emission is very faint and normalised out by
filtering (i.e. the background sky level is set to zero). To account
for this missing flux, we use the PACS aperture corrections of 0.87,
0.88, and 0.83 for 70, 100 and 160 $\mu$m (at 30, 30, and 26\arcsec).

The raw fluxes measured in each aperture are 472, 616, and 553 mJy,
and the NE background source estimates are 4.7, 11.8, and 15.3 mJy for 70,
100 and 160 \micron, respectively. Therefore, after subtracting the
background sources and dividing by the aperture corrections, the
fluxes are 537 and 687 mJy at 70 and 100 \micron, as shown in Table
\ref{fluxtable}, and 648 mJy at 160 \micron.  (We discuss the 160 \micron\ entry to Table
\ref{fluxtable} in the next paragraph.) The PACS observations are
fairly deep, so the uncertainties on these measurements are dominated
by calibration (2.64, 2.75, and 4.15\% at 70, 100 and 160 $\mu$m
respectively).  We therefore adopt uncertainties of 15 and 20 mJy at
70 \micron\ and 100 \micron, respectively.  Given these uncertainties
the flux densities of the disk are detected at a $S/N$ of 35 at both
wavelengths.

\begin{deluxetable}{rccc}
\tablecaption{Measured fluxes}
\tablewidth{0pt}
\tablehead{
\colhead{$\lambda$} & \colhead{$F_{aper}$\tablenotemark{a}} & \colhead{$F_{phot}$} & \colhead{$F_{disk}$} \\ 
\colhead{[\micron]} & \colhead{[mJy]} & \colhead{[mJy]} & \colhead{[mJy]} }
\startdata
70 & $537 \pm 15$ & $6.85 \pm 0.07$ & $533 \pm 15$ \\
100 & $687 \pm 20$ & $3.39 \pm 0.04$ & $712 \pm 20$ \\
160 & $570 \pm 50$ & $1.35 \pm 0.01$ & $569 \pm 50$ \\ 
250 & $309 \pm 30$ & $0.550 \pm 0.006$ & $307 \pm 30$ \\ 
350 & $163 \pm 30$ & $0.276 \pm 0.003$ & $163 \pm 30$ \\
500 & $74 \pm 30$ & $0.135 \pm 0.002$ & $74 \pm 30$ 
\enddata
\tablenotetext{a}{BG source subtracted through PSF fitting.}
\label{fluxtable}
\end{deluxetable}

In modelling the observations below, we find that the flux density
uncertainty is dominated by the background level at 160 \micron\ and
longer wavelengths.  The flux density increases with aperture size,
suggesting that the 650 mJy flux density likely includes some of the
extended background.  For example, a 30\arcsec\ aperture gives about
700 mJy, but a 40\arcsec\ aperture yields 760 mJy.  Obviously, not all
``estimates'' are equally good.  Given these uncertainties, we take
the model flux of 570 mJy as our ``best'' 160 $\mu$m value (as shown
in Table \ref{fluxtable}), but assign an uncertainty of 50 mJy (i.e.,
a $S/N$ of 11 is achieved on the disk flux density). This value is
higher than, but consistent with, the 539 mJy flux derived from {\it
Spitzer} MIPS at the same wavelength \citep{su09}.

At SPIRE wavelengths, the background dominates the images so the
photometry is less certain. The fluxes were estimated by subtracting
smooth models such that the background looked even, and adopting
conservative uncertainties. The \emph{Herschel} photometry and
uncertainties are presented in Table \ref{fluxtable}. The disk is well
detected at 250 and 350 \micron\ ($S/N$ of 10 and 5, respectively),
but has a flux detection of just 2.5 $\sigma$ at 500 \micron, where a
significant part of the total detected flux density arises from the
underlying cloud.

\subsection{Disk Size and Orientation}

To measure the disk size directly from the images, we did 2D
Gaussian fits to the star+disk emission of HR 8799.  We do not attempt
to fit the background source simultaneously with the star and
disk. The fits are first done to restricted areas around the target
due to the increasing cirrus emission at longer wavelengths. After the
full width half maximum and position angle of the fits is established,
the models are extended to larger scales. 
The orientation of the best-fit Gaussians changes between the PACS and
SPIRE data; it is likely that the background source (unresolved at these wavelengths from the HR 8799 disk), is impacting the orientation of the fit in these data.  The sizes of the fitted
Gaussians derived from the PACS maps are shown in Table
\ref{tbl:gaussianFits}. These fits provide consistent results for
the 70 and 100 \micron\ data, both for the deconvolved size of the
FWHM of the disk and its inclination and position angle.

\begin{deluxetable*}{rccccccccc}
\tabletypesize{\footnotesize}
\tablewidth{0pc}
\tablecaption{\label{tbl:gaussianFits} 2D Gaussian fits to star+disk emission in PACS maps }
\tablehead{
\colhead{Band} & \colhead{beam} & \multicolumn{2}{c}{FWHMs} & \multicolumn{2}{c}{Deconvolved} & \colhead{Disk Size\tablenotemark{a}} & \colhead{Inclination\tablenotemark{b}} & \colhead{Position Angle} \\ 
\colhead{($\mu$m)} & \colhead{(\arcsec)} & \colhead{major (\arcsec)} & \colhead{minor (\arcsec)}  & \colhead{major (\arcsec)} & \colhead{minor (\arcsec)} & \colhead{(AU)} & \colhead{(\degr)} & \colhead{(\degr\ E of N)} }  
\startdata
70 & 5.61 & 15.3 $\pm$ 0.2 & 13  $\pm$ 0.2 & 14.2 $\pm$ 0.2 & 12.7 $\pm$ 0.2 	& 567 $\pm$ 8 & 26.2 $\pm$ 3.3 	& 62.1 $\pm$ 5.6 \\ 
100 & 6.79 & 16.4 $\pm$ 0.3 & 15.2 $\pm$ 0.3 & 15.0 $\pm$ 0.3 & 13.6 $\pm$ 0.3 	& 599 $\pm$ 12 & 24.6 $\pm$ 5.1 & 62.9 $\pm$ 4.1 \\ 
160 & 11.36 & 18.9 $\pm$ 0.4 & 17.0 $\pm$ 0.3 & 15.1 $\pm$ 0.5 & 12.6 $\pm$ 0.4 & 602 $\pm$ 20 & 33.4 $\pm$ 6.5 & 54.8 $\pm$ 5.3 
\enddata
\tablenotetext{a}{Adopting a distance of 39.9 pc for HR 8799 and assuming a circular disk.}
\tablenotetext{b}{Calculated using deconvolved disk sizes.}
\label{gaussfits}
\end{deluxetable*}

\subsection{Spectral Energy Distribution}
\label{sed}

\begin{figure}
\begin{center}
\hspace{-0.5cm} \includegraphics[width=0.5\textwidth]{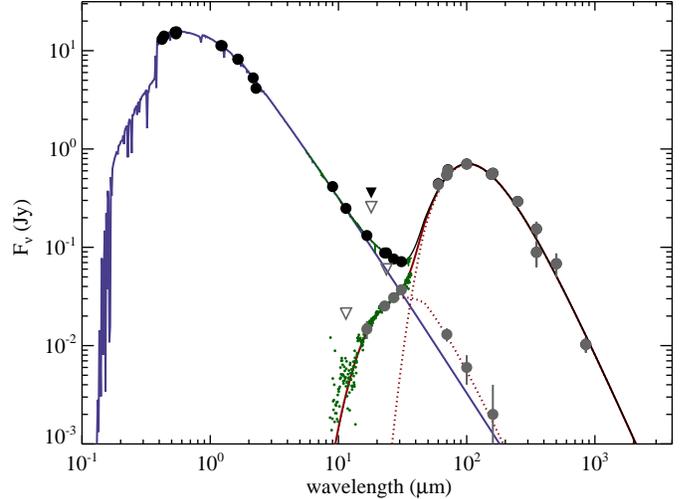}
\caption{Star + disk SED for HR 8799, showing flux densities (dots) and upper limits (triangles). Shown 
are observed (black symbols) and star-subtracted fluxes (grey
symbols), and the IRS spectrum (green dots). At long wavelengths, the gray dots cover the black dots completely, due to the declining flux density of the photosphere.  At intermediate wavelengths, the star-subtracted flux densities are divided between the warm and cold disk components; therefore, there are two flux densities plotted for those wavelengths. The stellar photosphere
model is shown as a blue line, the disk model as a red dotted line,
and the star+disk spectrum as a black line.}
\label{fig:sed}
\end{center}
\end{figure}

Figure \ref{fig:sed} shows the spectral energy distribution (SED) for
HR 8799 using photometry from a wide range of sources. For the stellar
component we fit PHOENIX models
\citep{Bro05} to optical and near-IR photometry
\citep{syl96,per97,hog00,cut03}, finding $T_{\rm eff}=7380$ K and $L_\star=5.4 L_\odot$, 
equivalent to a late-A or early-F spectral type star, consistent with HR 8799's published spectral 
types \citep[A5V, F0V,][]{gra99,gra03}.

The \emph{Herschel} photometry from 70-500 $\mu$m provides coverage
over a wide range, and combined with previous data
\citep{mos90,wil06,su09,ish10,pat11},
gives a fairly complete picture of the disk spectrum. To this
photometry we have fitted a two-component modified blackbody
model. The ``modification'' refers to two additional parameters,
called $\lambda_0$ and $\beta$, with which a pure blackbody spectrum
is modified beyond $\lambda_0$ by the multiplicative factor $\left(
\lambda_0/\lambda \right)^\beta$. This factor attempts to account for inefficient emission by grains that are small relative to the wavelength of their emission. Therefore, $\lambda_0$
should in some way be representative of the grain size that dominates
the emission spectrum.  HR 8799 is also thought to have an unresolved
inner warm disk component. {\it Herschel} photometry for this
component is uncertain due to degeneracy with parameters for the outer
disk, see \S \ref{ss:models} for further discussion.

The best fit model shown in Figure \ref{fig:sed} has temperatures of
$153 \pm 15$ K and $36 \pm 1$ K for the two dust components, with
$\lambda_0=47 \pm 30$ \micron\ (upper limit 90 \micron) and $\beta=1.0
\pm 0.1$ common to both. Compared to most known debris disks, where
$\lambda_0$ is usually 100 \micron\ or longer \citep[e.g.,][]{boo13},
the relatively short $\lambda_0$ here suggests that the disk spectrum
is dominated by smaller grains than is typical in other debris
disks. Given that HR 8799 is already known to be somewhat unusual due
to its very extended disk that may comprise small, radiation-dominated
grains on eccentric orbits, is seems likely that the short $\lambda_0$
is somehow related to the halo.

\section{Image Modelling}\label{ss:models}

\subsection{\emph{Herschel}}

Our goal here is to present a simple characterisation of the \emph{Herschel} observations
of HR 8799, not a detailed model that includes possible grain properties. We therefore
model the data at each wavelength independently to arrive at a surface brightness profile
for each wavelength. For the SPIRE data, which are strongly affected by the background
level, we scale the PACS models to make the background level look smooth to derive
disk+star flux densities.

The parameterized model is basically the same as used previously for
modelling \emph{Herschel} PACS observations
\citep{ken12a,ken12b,wya12}. Because we
model each wavelength independently, however, the temperature profile
is not needed and the model produces surface brightness images to
compare with the data. The models are inspired by the \citet{su09}
model, comprising three distinct components. The first is unresolved,
which we simply add to the stellar flux when generating the model
images. The other two components are called the ``belt'' and
``halo'', extending from $r_1$ to $r_2$ and $r_2$ to $r_3$. Each
wavelength has a power-law change in surface brightness with radius
for each component (i.e. $\alpha_{\rm \lambda,comp}$). As found by
\citet{su09}, the belt component has a reasonably constant surface
brightness, while the halo component drops steeply with radius.

The PACS models are created as three dimensional structures and then
viewed with a specific geometry specified by the disk inclination and
position angle to generate an image. These images are generated at
high resolution and then convolved with the PACS beam (calibration
observations of $\alpha$ Boo processed in the same way as the HR 8799
data), resulting in an image that can be directly compared with the
science data. The overall brightness of each component and the
power-law indices were varied to find a sufficiently good match to the
data at each wavelength. We also varied $r_1$, $r_2$, and $r_3$,
though these were required to be the same at all three wavelengths so
the models are not completely wavelength independent.

A final component for the PACS models is the NE background source,
which is simply added as a background point source whose location is
the same at each wavelength, but whose brightness is allowed to vary
with wavelength.

For SPIRE, we used a model that attempted to achieve an overall global
fit to all data simultaneously rather than choose one of the
individual PACS models. The resolution at these wavelengths is not
sufficiently good, and the background level sufficiently high, that
this choice is not important, given that these models were only used
to provide estimates of the disk photometry.

\begin{figure*}
\vspace*{15cm}
\includegraphics{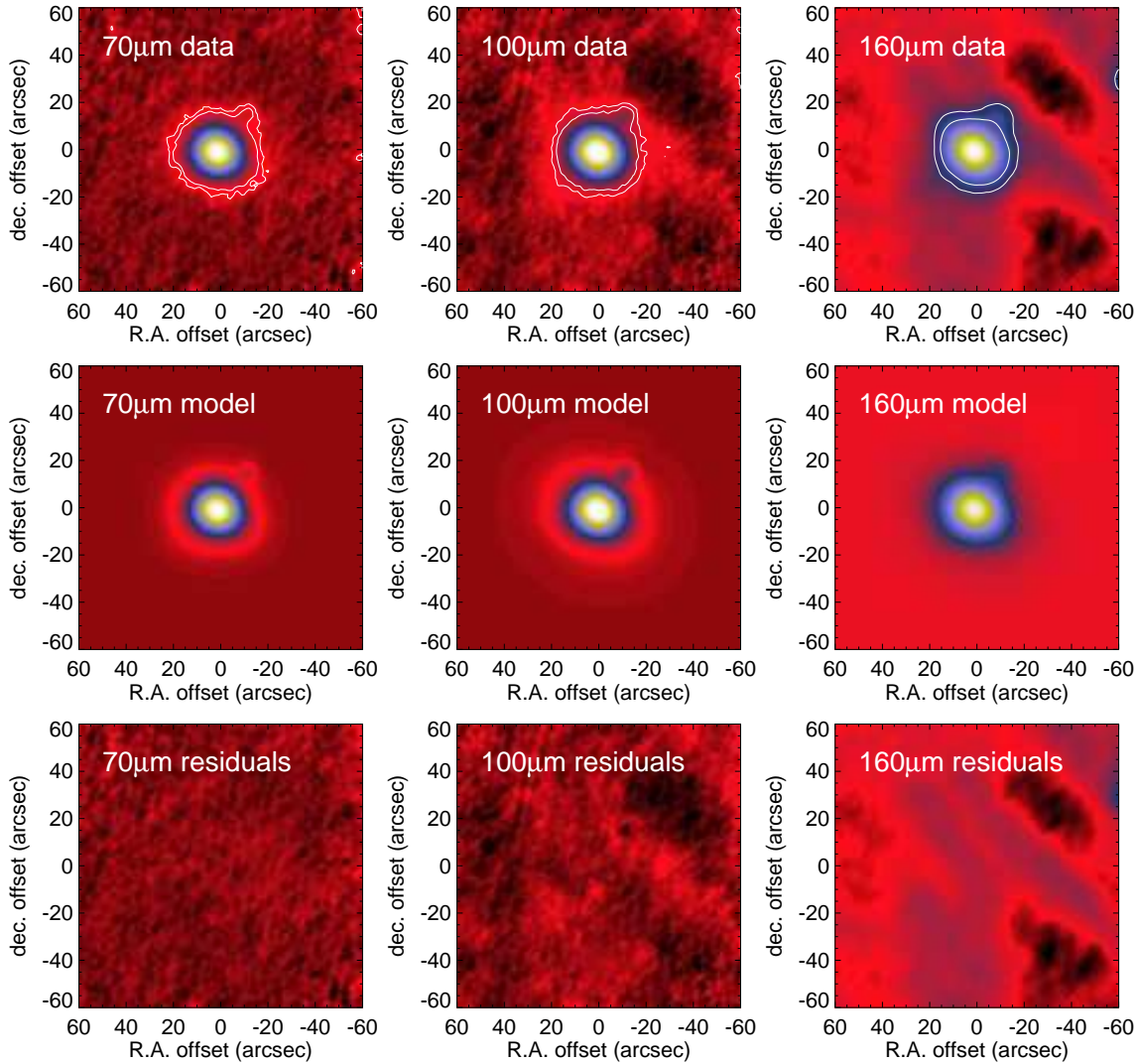}
\caption{Results of PACS data models, showing the PACS data, model for each wavelength and the residual. 
}
\label{modelimages}
\end{figure*}

The results of the modelling are shown in Figure \ref{modelimages},
which shows the PACS data, each model, and the residuals after
subtracting the model. While some residual structure remains due to
variations in the background level, in each case there is no evidence
for significant departures from azimuthal symmetry. We return to this
topic in the Discussion section ($\S$ \ref{discussion}).

\begin{figure}
  \begin{center}
    \hspace{-0.5cm} \includegraphics[width=0.5\textwidth]{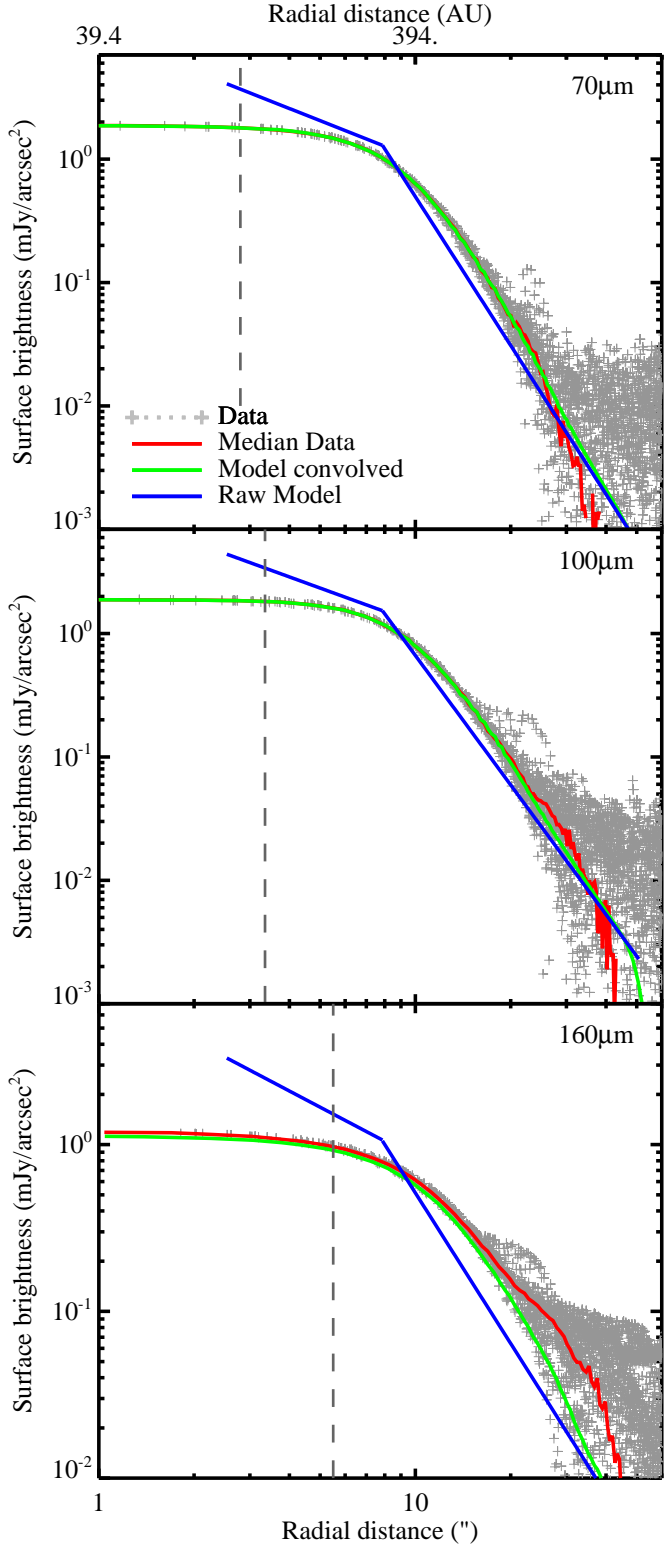}
    \caption{HR 8799 radial profiles at the three PACS wavelengths. Each panel shows the
      raw surface brightness values for all pixels within 60\arcsec\ of the star (dots) and
      their median (red), the median convolved model surface brightness (green), and the
      raw unconvolved profile (blue). All profiles are generated using the image models,
      and assume a disk inclination of 25.75$^\circ$ and PA of 63.73$^\circ$. The dashed grey lines represent the half-width at half-maximum of the beam.}
\label{fig:prof}
  \end{center}
\end{figure}

We found radii of $r_1=100$ AU, $r_2=310$ AU, and $r_3=2000$
AU.  The best fit power-law parameters
and unresolved component fluxes are given in Table \ref{tab:model}. We
do not consider the contraints on the unresolved flux, the belt inner
edge, and the belt power-law index to be very strong, particularly at
160 \micron; the belt inner edge lies within the PACS beam so these
parameters are all degenerate (see also Fig
\ref{fig:prof}). At 160 \micron, we have in fact fixed the unresolved flux and the belt and
halo power-law indices to ``sensible'' values, by which we mean they
do not lead to unusual conclusions about the radial temperature
structure in the outer disk (see below). The outer disk radius is also
poorly constrained and could continue to larger radii where the
emission is here dominated by noise. The outer radius is considered representative to within a few hundred AU, but we note the caveat that we cannot be assured that all the flux has been captured in our maps, given the filtering and PSF effects (see $\S$ \ref{datared}). The power-law indices at 70 and
100 \micron\ for the halo are moderately well constrained (under the
assumption that a power law is correct), with values of $\alpha_{\rm
Halo} = -4 \pm 0.3$ and $-3.5 \pm 0.5$ at 70 and 100 \micron\
respectively, with the uncertainties derived from the radial
profiles fits, rather than the model images. The value of 13 mJy for the unresolved
component at 70 \micron\ agrees reasonably well with the predicted
value of about 18 mJy if this component is a pure blackbody
\citep{su09}. The values of 6 and 2 mJy at 100 and 160 \micron\ are
less certain due to poorer resolution, but these are less than
expected for a pure blackbody based on the 13 mJy value at 70
\micron. However, given that all three lie along a line of $F_\nu
\propto \lambda^{-2}$, and that this dependence is steeper than
expected for a 150 K blackbody, it seems likely that the emission spectrum
of the warm component is not a pure blackbody. Such a conclusion would
be unsurprising given that debris disk spectra are commonly seen to be
steeper than a blackbody in the far-IR. That the turn-over wavelength
($\lambda_0$) is shorter than usually seen may be indicating
relatively small grains or a relatively steep size distribution. We discuss this further in $\S$ \ref{smallgrains}.

The disk inclination is 26$^\circ$, for which we consider the
uncertainty to be about 3$^\circ$. The results from image modeling are
completely consistent with the inclination based on 2D Gaussian fits
to the data, and so do not appear to be strongly model dependent.  To
account for correlated noise in the images, the uncertainty in the quantities derived from the images -- inclination and position angle -- have been
multiplied by 3.6 based on \cite[][see \cite{ken12a} for more
details.]{fru02}.  To estimate this uncertainty formally would be
difficult as many model parameters would need to be marginalized
over. To check that the estimate of a few degrees is sensible, we
fitted 2D Gaussians to a star-subtracted image to see how the fit
quality varied, with the same conclusion that the uncertainty is a few
degrees. The position angle of the disk is $62 \pm 3^\circ$ East of
North.

\begin{deluxetable}{rrll}
\tablecaption{Wavelength-dependent model parameters.}
\tablewidth{0pt}
\tablehead{
\colhead{$\lambda$ [$\mu$m]} & \colhead{$F_{\rm unres}$} [mJy] & \colhead{$\alpha_{\rm Belt}$} & \colhead{$\alpha_{\rm Halo}$} }
\startdata
    70 & 13 & -1.0 & -4.0 \\
    100 & 6 & -0.9 & -3.5 \\
    160 & 2 & -1.0 & -3.0
\enddata
\tablecomments{The parameters 
    are poorly constrained at 160 \micron\ because the inner part of the disk is poorly
    resolved ($F_{\rm unres}$, $\alpha_{\rm belt}$, and $r_{1}$ are degenerate) and the
    background is much higher ($\alpha_{\rm halo}$ is uncertain). The values at 160 \micron\
    were therefore chosen to agree with the conclusions based on 70-100 \micron\ data (see text).
}
\label{tab:model}
\end{deluxetable}

As a check on the aperture photometry in \S \ref{ss:phot}, the total
model fluxes, including the star but not the background source, are
summarized in Table \ref{tab:modelfluxes} (see photometry discussion
above regarding the 160 \micron\ flux). Splitting these into
contributions from the belt and halo components is also informative,
given that the conclusion from the SED was that small grains dominate
the overall spectrum. These component fluxes are also shown in Table
\ref{tab:modelfluxes}.  Therefore, the halo component contributes a 
significant amount of the flux at all {\it Herschel} wavelengths and 
dominates beyond 160 \micron. If the halo comprises mostly small grains 
then the relatively short $\lambda_0$ value is not surprising.

\begin{deluxetable}{rccc}
\tablecaption{Relative Flux Contributions from the Belt and Halo.}
\tablewidth{0pt}
\tablehead{
\colhead{$\lambda$ [$\mu$m]} & \colhead{$F_{\rm model}$ [mJy]} & \colhead{$F_{\rm belt}$ [mJy]} & \colhead{$F_{\rm halo}$ [mJy]} }
\startdata
70 & 550 & 304 & 215 \\
100 & 700 & 339 & 325 \\
160 & 570 & 236 & 298 
\enddata
\label{tab:modelfluxes}
\end{deluxetable}

Because the HR 8799 disk appears azimuthally symmetric, we now
consider the radial profiles of both the observed and modelled
emission. These are shown in Figure \ref{fig:prof}, where it is again
clear that the model reproduces the data well. The break in the
surface brightness profile around 10\arcsec\ is apparent, and
justifies use of the two-component belt+halo model for the outer
disk. As noted above, the steepness with which the belt surface
brightness drops with radius is somewhat degenerate with the
unresolved flux from the warm component, particularly at 160
\micron. While the model drops below the observed profile at this
wavelength, it leaves a smooth background so is satisfactory. It is
also clear that the data constrain the surface brightness profile
fairly well out to 20-30\arcsec\ (800-1200 AU) while the emission is
detected out to 2000 AU, particularly at 100 and 160 \micron.

\subsection{\emph{Spitzer} MIPS 24$\mu$m}

\begin{figure}
  \begin{center}
    \hspace{-0.5cm} \includegraphics[width=0.5\textwidth]{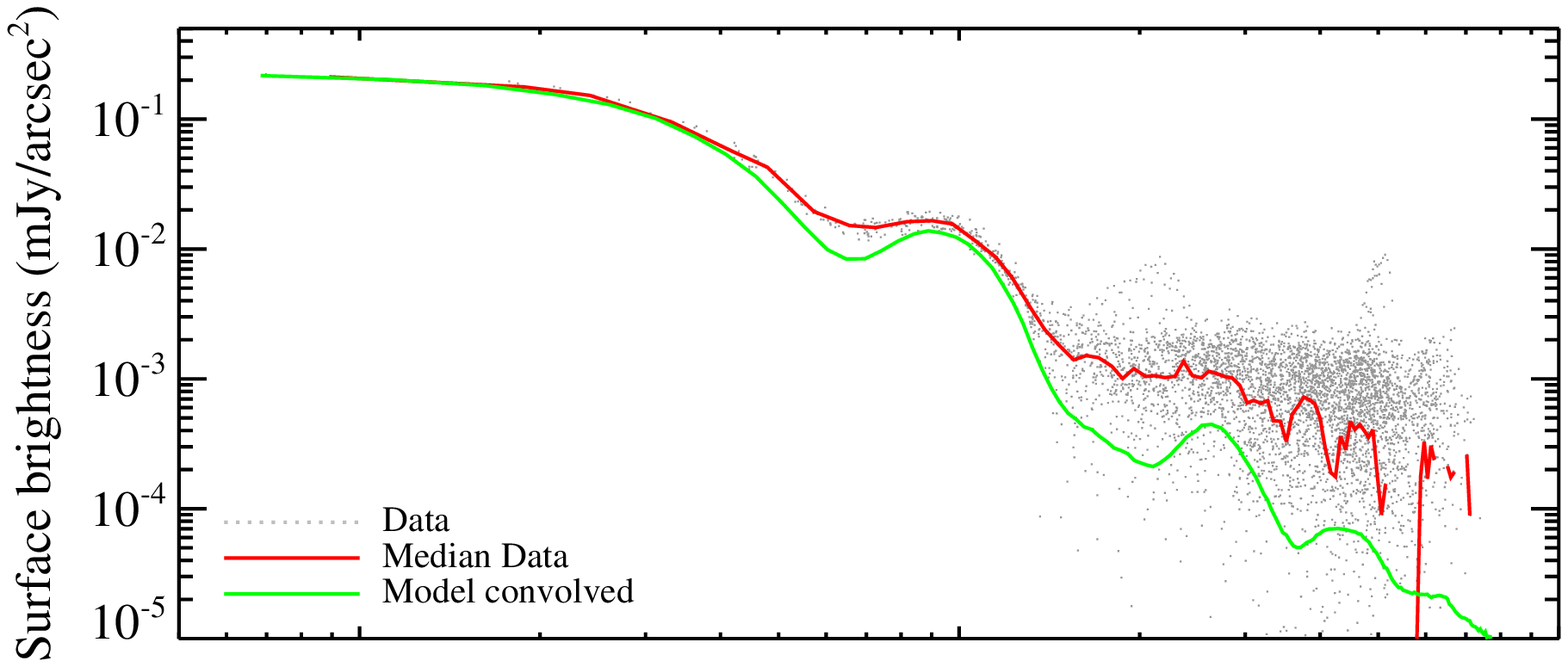}\\
    \hspace{-0.5cm} \includegraphics[width=0.5\textwidth]{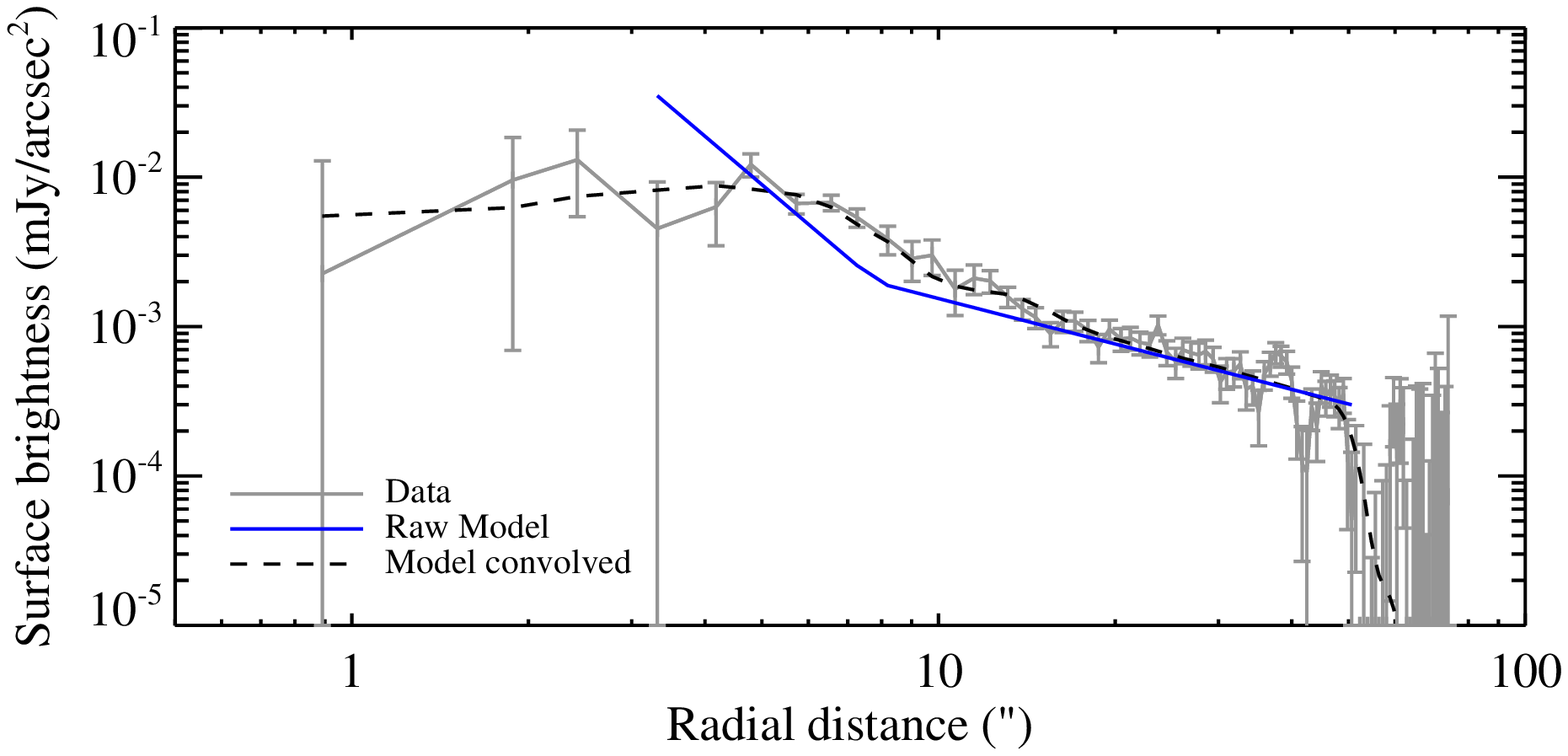}
    \caption{\emph{Top:} Radial profile at 24 \micron. \emph{Bottom:} 
      Star-subtracted outer
      disk profile and surface brightness model fit. Aside from a difference in units,
      panels are directly comparable with Figure 6 of \citet{su09}.}
\label{fig:prof24}
  \end{center}
\end{figure}

While the PACS data are reasonably well resolved, they do not provide particularly wide
coverage in terms of wavelength. Therefore, we now consider the disk as resolved at
24 \micron\ by \emph{Spitzer} \citep{su09}. Our starting point is a
star-subtracted 24 \micron\ image processed as described in that paper. The extended disk
emission in this image has relatively low S/N, so we only consider the radial profile,
which is shown in the top panel of Figure \ref{fig:prof24}.

To derive the profile of the extended disk component we follow
\citet{su09} and subtract a point source model scaled to the peak emission
at the stellar position (here a 3.5\arcsec\ boxcar-smoothed STinyTim model) to remove the
contribution from the unresolved warm disk emission. This procedure is justified because
the total belt+halo emission is at most $\sim$5\% of the total 24 \micron\ emission, of
which only a small fraction is expected to arise from the stellar position. The resulting
belt+halo radial profile is shown in the bottom panel of Figure \ref{fig:prof24}.

As described by \citet{su09}, the disk is seen to be extended. The
significance of the signal is weak near the stellar position due to only being slightly
brighter than the point source model, and weak at large radii due to low S/N. However,
the signal is strong from 5-50\arcsec, where it decays monotonically. A small difference here
is that we find that the surface brightness is significantly above zero to slightly
larger distances (about 60\arcsec\ rather than 50\arcsec). The overall trend is similar, in that the
surface brightness decays from about $10^{-5}$Jy arcsec$^{-2}$ at a few arcseconds radius
to a few $10^{-7}$Jy arcsec$^{-2}$ at 50\arcsec\ radius.

To derive the underlying surface brightness distribution, we convolved a two-power-law
model (i.e. belt+halo) with the STinyTim PSF to find the best fitting via $\chi^2$
minimisation. During this process we fixed the belt and halo radii to those found from
the PACS data (i.e. 100, 310, and 2000 AU). The best fitting model is shown in the bottom
panel of Figure \ref{fig:prof24}. While it is clear that the slope of the halo component
is fairly well constrained, the inner component is not due to the large surface
brightness uncertainty at small radii. The power-law slopes are $-3.3 \pm 0.1$ and $-1
\pm 0.06$ for the belt and the halo, respectively, with the uncertainties derived from the diagonal covariance matrix
elements. The uncertainties are probably somewhat larger due to systematic uncertainties,
such as the correctness of the point source model. However, as we show below, the observed
profile is interesting when compared with the \emph{Herschel} data.

\subsection{Combined radial profiles}

\begin{figure}
  \begin{center}
    \hspace{-0.5cm} \includegraphics[width=0.5\textwidth]{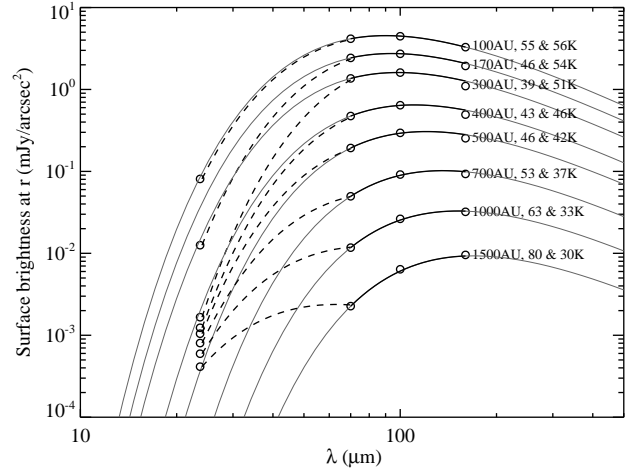}
    \caption{Disk emission profile at different radii. A blackbody fit to the
      70-100$\mu$m data is shown at each radius (solid lines) as well as a blackbody fit
      to the 24-70$\mu$m data (dotted lines), with the two temperatures given at each
      radius. With only three reliable wavelength points at each radius the temperatures
      are degenerate, so here we just show independent fits to each pair of 24-70 and
      70-100$\mu$m.}\label{fig:spec}
  \end{center}
\end{figure}

Though the surface brightness profiles highlight the radial structure well, the
unconvolved model profiles can be taken one step further to show the disk emission
spectrum as a function of radius, as shown in Figure \ref{fig:spec}. These combine the
MIPS and PACS data described above. At each radius we have divided by a colour correction
calculated using a blackbody given by the 70-100 \micron\ flux ratio for the 70, 100 and 160$\mu$m
fluxes, and a correction using the 24-70 \micron\ correction for the 24 \micron\ fluxes. This
correction is only a few percent in the PACS bands and has a maximum of 1.3 at 300 AU
(39 K) for MIPS. As noted below the temperatures used for these colour corrections are
uncertain, but as they are small, different temperatures will not significantly change our conclusions.

For all radii Figure \ref{fig:spec} shows that the disk spectrum at the PACS wavelengths
of 70, 100 and 160 \micron\ is well described by a blackbody. However, this agreement is contrived at
160 \micron\ due to uncertain model parameters. Therefore, we have only used the
$70+100$ \micron\ data to derive the blackbody temperature at each wavelength. We also derive
a blackbody temperature for the $24+70$ \micron\ colour. While we would ideally fit two
blackbodies simultaneously, with only three wavelength measurements (ignoring 160 \micron)
at each radius the temperatures are degenerate so we take these independent values as
representative.

The main point is that the derived temperatures are significantly different for $24+70$ versus
$70+100$ \micron. At the innermost distances the agreement is reasonable given the
uncertainty in the MIPS profile, but not beyond $\sim 500$ AU. The relatively flat
power-law decay of the MIPS surface brightness profile means that as the disk temperature
as derived from PACS drops, the observed MIPS surface brightness becomes increasing high
compared to the expected level. This discrepancy might be questioned at the outermost
radii, where small differences in the assumed background level could cause the MIPS
profile to be lower (e.g., zero at 50\arcsec). However, at 20\arcsec\ (800 AU) the discrepancy is
harder to explain away. The expected 37 K disk temperature predicts a 24 \micron\ surface
brightness of about $10^{-5}$ mJy arcsec$^{-2}$, but the observed level is two orders of
magnitude higher. The radial profiles from \citet{su09} are slightly
fainter than ours, but the discrepancy is still far larger than expected.

To achieve even a rough agreement of the 24 \micron\ radial profile with the levels
predicted by the PACS 70-100$\mu$m temperatures requires subtracting a DC level of $7
\times 10^{-4}$ mJy arcsec$^{-2}$ from the MIPS image. The median and mean pixel levels are
$-1 \times 10^{-4}$ and $-6 \times 10^{-5}$ mJy arcsec$^{-2}$ respectively, with a standard
deviation of $10^{-3}$ mJy arcsec$^{-2}$. Therefore, such a subtraction is unjustified
because it shifts the mean value in the MIPS image by nearly 1$\sigma$. Further, such a
subtraction causes the median radial profile to become negative beyond 30\arcsec\ (i.e. in a
region where Figure \ref{fig:prof24} shows it is still clearly decreasing with radius).

Therefore, while there is some uncertainty in the absolute level, it seems that the
levels predicted from the temperatures derived from the PACS data and the levels seen by
MIPS are in clear disagreement, most significantly at large radii ($\gtrsim 500$ AU). The
same conclusion would be reached using the radial profiles derived by
\citet{su09}, where there is still significant disk emission at 30\arcsec\ (a
few $10^{-4}$ mJy arcsec$^{-2}$ at 1200 AU), where the PACS-predicted level is less than $10^{-5}$ mJy arcsec$^{-2}$, as indicated in Figure \ref{fig:spec}. Indeed,
their favoured model had a halo component that was warmer than the belt component,
despite the larger distance, which was attributed to the halo comprising smaller grains
and seems consistent with the $24+70$ \micron\ temperature profile derived here.

Alternatively, the PACS 70 or 100 \micron\ profiles would need to be at least a factor of
two in error at these distances to allow a single temperature consistent with the MIPS
data. However, the good agreement between the data and model in Figure \ref{fig:prof}
show that such a difference is very unlikely.

\section{Discussion}
\label{discussion}

\subsection{Evidence for Azimuthal Structure}
\label{asymmetric}

Visual inspection of the data reveals no apparent asymmetries
in the emission from the HR 8799 disk, excepting the compact northwest
feature.  This is in contrast to recent findings in 350 \micron\ maps
from the CSO \citep{pat11}.  Non-axisymmetric structure was observed,
which was interpreted as evidence of particles trapped in a 2:1
resonance with the outer planet. We note that our resolution at 350
\micron\ is poorer than that of the CSO by an approximate factor of
2. Other than those ground-based data, no asymmetries have been seen
in the disk. \cite{hug11} attribute the asymmetries in interferometric
maps from the SMA to noise features in low signal-to-noise data.  Our
data are consistent with this picture. We note that strong asymmetries
are not typically expected at far-infrared wavelengths, but rather
models suggest that the strongest asymmetries will be observable in
the submillimeter where the data sample the larger grains in the disk
which are more likely to be localized in the planetesimal belts and
trapped in resonances.  

To place a limit on the brightness of a clump which would have been
detected, we take the 5$\sigma$ limit at 70 \micron\ as an
example. The rms at 70 \micron\ is 0.34 mJy beam$^{-1}$, and to see a
believable substructure in the disk would have required a peak flux of
1.7 mJy.  In contrast, the symmetrically distributed disk emission has
a 70 \micron\ flux of roughly 530 mJy. Therefore, a ``discrete'' peak
in the disk of flux density $\sim 0.3$\% of the symmetrical emission
should have been detectable.  The 70 \micron\ beam has a scale of 220
AU at the distance of HR 8799, so even unresolved point sources could
cover a significant fraction of the inner disk.  Our ability to place
limits on our sensitivity to still larger resolved clumps in the disk
is degraded by the large filter scale applied during data reduction, a
scale required to accurately recover the largest scales of the halo.

\subsection{Implications for Planet Formation Scenarios}

The fact that no significant azimuthal structures were observed in the
Herschel images of the HR 8799 disk is significant because a planet
formation scenario in which planets form close to the star and then
migrate outward to their present positions would have favored the
trapping of planetesimals into resonances \citep{wya03}. The resulting
resonant structure would result in the planetesimal distribution
having a clumpy structure, and observations of the level and
morphology of that clumpiness would be indicative of the planets'
migration rate and therefore of the mechanism which caused the
migration \citep[e.g.,][]{cri09}. Dynamical stability studies have
suggested that the HR 8799 planets could themselves be in resonance
\citep{fab10,rei09,goz09,goz13}.  This does not de facto imply
migration has occurred in the system, since the formation of planets
in situ by gravitational instability \citep[e.g.,][]{dod10} can also
result in resonant configurations.

Comparison with observations of the disk is complicated by the fact
that we observe the distribution of dust, and not of
planetesimals. Observations at wavelengths $>200$ \micron\ are
sensitive to dust that is large enough to trace the planetesimal
distribution. However, shorter wavelength observations are sensitive
to small dust sizes that can have an axisymmetric distribution even if
the planetesimals are themselves in resonance \citep{wya06}. Given the
low resolution of our SPIRE data, and the short wavelength of the
higher resolution PACS images, we cannot set strong constraints on the
resonant structure of the planetesimal belt in this system from our
{\it Herschel} observations, though this is implied by the previous
publication of an asymmetry in the disk at 350 \micron\
\citep{pat11}. If such asymmetric structures are present, then they
should be most readily detectable by ALMA.

\subsection{Constraints on System Inclination and Orientation}

From our disk model based on well-resolved images, the inclination of the
disk is $26 \pm 3$\degr.  Given the azimuthal
symmetry we observe, it is unlikely that the disk is inclined at
significantly $> 25$\degr.  Similarly, the \cite{su09} {\it Spitzer} data
also ruled out inclinations larger than $\sim 25$\degr.  Inclinations
higher than 45\degr\ are not supported by the disk images, assuming a
circular disk geometry inherent in the disk system.  

Estimates of the stellar inclination from astroseismology suggest $i >
40$\degr\ \citep{wri11}, although a value as low as 35\degr\ is not
ruled out by their work (see their Figure 3), but these values are
inconsistent with the modeling of
\cite{rei09} which suggest a near face-on inclination of
13-30$^\circ$. This is consistent with the orbit modeling for the
detected planets \citep{mar08,fab10,laf09,sou11}, and with estimates
from the debris disk modeling presented in \cite{su09}.

The suggestion that the entire HR 8799 system may be coplanar is
consistent with the results of \cite{wat11} and \cite{gre13}, which includes
HR 8799, who found common inclinations between a significant
number of stars for which the stellar and disk inclinations had been
independently measured. In addition, the recent work by \cite{goz13}
on the astrometry of the HR 8799 system reports an inclination of
$\sim 24$\degr\ for the planetary orbits and measures the longitude of
the ascending node to be $\sim 64$\degr.  The disk position angle ($62
\pm 3$\degr) is measuring the disk longitude of ascending node, so if
the planets are coplanar with the disk, we should expect both values
to agree within uncertainties, which they clearly do.  It is therefore
likely that the entire HR 8799 star-disk-planet system is coplanar.

\subsection{Constraints on the Eccentricity of the Disk}

Any eccentricity in the disk is degenerate with the
inclination; therefore, the best constraint on non-axisymmetry comes from the
fact that one side is not seen to be brighter than the other. 
Conservatively, the PACS 70 \micron\
observations are sensitive to a $\le $1\% brightness asymmetry, which
corresponds to roughly a 2\% difference in dust radial distances from
one side of the disk to the other (assuming a linear dependence of
flux density on temperature).  Such an
asymmetry would arise if planetary eccentricities caused the disk to
become offset from the star due to secular perturbations (i.e as seen
for the debris ring around Fomalhaut, Kalas et al. 2005). Fomalhaut
shows a clear asymmetry from PACS images of similar S/N with an
eccentricity of 0.1 (Acke et al. 2012).  The inner edge of the cold
belt of HR 8799's disk is at a similar radial distance as that of
Fomalhaut; therefore, the disk eccentricity expected for HR 8799 must
be comparable to, or below, that level.

\subsection{Preponderance of Small Grains}
\label{smallgrains}

The relatively short value of $\lambda_0 < 90\mu$m for the outer
component (belt+halo) from the SED (\S \ref{sed}) suggests that the
emission from this component is dominated by grains smaller than this
wavelength (or equivalently has a relatively steep size distribution
overall). It seems likely that this dominance is related to the
presence of the outer halo, though interestingly the disks around
Fomalhaut, Vega and $\beta$ Pictoris, the best studied systems with
haloes, have $\lambda_0$ from 150-250$\mu$m. Clearly then, the
existence of a detectable halo does not require a short
$\lambda_0$. The question may therefore be one of their relative
brightness; a brighter halo weights the overall size distribution to
small sizes. One way to increase the halo brightness is with a
recently increased level of dust due to collisions, this increase both
injects more particles into the bound (high eccentricity) part of the
halo, as well as the unbound (para/hyper-bolic) part. Little work has
been done to compare halo models with those observed, most notably
\citet{mul10} found that the halo around Vega could be
explained with a steady-state halo model, in contrast to the original
suggestion that it was due to a recent event
\citep{su05}. The much shorter value of $\lambda_0$
seen for HR 8799 suggests that a similar analysis should be done, with
a particular focus on the possible origins of an overabundance of small
($<$50 $\mu$m) grains compared to the Vega disk.

Similarly, the existence of 24 $\mu$m emission at larger radii than
expected (Fig.\ \ref{fig:prof24}) may be related to the presence of small
grains. However, in this case the reason is unclear. A possible cause
is that grains in the halo have more than one composition, thus
resulting in more than one temperature at a single radial
distance (although the narrow SED would seem to rule this out). 
Another potential reason is that HR 8799 is passing through
an overdense part of the ISM, and that at least some of the observed
SED actually comes from heated ISM dust rather than circumstellar
dust. There is however no evidence in the images that HR 8799 is
strongly perturbing the ISM around it, as has been observed for the
nearby star $\delta$ Velorum \citep{gas08}.

The temperature of the blackbody fit implies a disc radius of 140 AU, 
which lies within the true radial distribution of the dust. This is 
surprising as the luminosity of HR 8799 ($5.4 L_\odot$) results in a low 
blowout grain size meaning that we would expect there to be plenty of 
small grains in the belt itself, and these would be much hotter than blackbody 
\citep{boo13}. Such an inconsistency is not unheard of; for 
instance, \cite{lohne12} found that, for HD 207129, the dominant 
grain size was ten times higher than the blowout limit. The lack of 
small grains in the belt does also place extra emphasis on the dust in 
the halo being the dominant component to the SED.

As discussed in $\S$
\ref{asymmetric}, based on the modeled symmetric disk, an unresolved
clump of flux density $< 1$\% of the total flux at 70 \micron\ would
have been detectable. The smoothness could indicate that the halo is
populated with bound high eccentricity small grains, as has been suggested
for Vega based on {\it Herschel} observations \citep{sib10}, since such grains would 
have an axisymmetric distribution regardless of the resonant structure of 
the planetesimal belt \citep{wya06}.  However, a population of unbound 
grains could also be axisymmetric, and moreover the evidence for clumping 
in the parent planetesimal population is limited.

\section{Summary}
\label{conc}

We have imaged the HR 8799 debris disk with PACS and SPIRE on {\it
Herschel} and modeled resolved imaging at 70, 100 and 160 \micron.
Based on the SED, we find evidence of two components, a warm
unresolved component of $153 \pm 15$ K and a cold, resolved component
of $36 \pm 1$ K.  Imaging modeling reveals two distinct components of
cold dust, as first identified by \cite{su09}. We find evidence of a
cold planetesimal belt, a ``Kuiper Belt analogue'', extending from
100-310 AU, and a more extended halo component from 310 - 2000 AU.
The radial profiles of the dust emission show a clear break at 310 AU
in the power-law slopes, from shallower slopes of $-0.9$ to $-1.0$ for
the planetesimal belt to steep slopes of $-3.0$ to $-4.0$ for the halo
(depending on wavelength).  

The radius inferred from the blackbody fit to the cold component of
the SED is 140 AU, well within the planetesimal belt of HR 8799's
disk. Therefore, unlike many other disks in which the grains are found
at considerably larger radii than their blackbody fits would predict
\citep{boo13,rod12}, in HR 8799 the grains appear to be very well fit
as blackbodies.

The fluxes from the planetesimal belt and halo are comparable. The SED
indicates a steepening of the spectrum at $\lambda_0 = 47 \pm 30$
\micron, with an upper limit of 90 \micron. Taken as a representative of 
the grain size that dominates the emission spectrum, it therefore
appears that HR 8799 is populated by smaller grains than other A star
disks with haloes (e.g., Vega, $\beta$ Pic and Fomalhaut), as well as
A star disks in general \citep{boo13}.

The combined radial profiles of our {\it Herschel} data and {\it
Spitzer} 24 \micron\ data \citep{su09} show that the derived
temperatures are significantly different for $24 + 70$ \micron\ versus
$70 + 100$ \micron\ beyond 500 AU (i.e., in the halo).  Therefore, while
the PACS data imply that the halo is colder than the planetesimal
disk (temperature decreasing with radial distance), the $24 + 70$
\micron\ data suggest that the halo is warmer, consistent with the
conclusions of \cite{su09}.  This could mean there are potentially two
distinct populations of dust grains (sizes or compositions) in the
halo and reinforces the importance of obtaining resolved imaging at
multiple wavelengths to properly understand the interior physical
structure of disks.

We constrain the inclination of the disk to be $26\pm3$\degr\ and the
position angle to be $62\pm3$\degr. These values are determined
through image modeling and also through 2D Gaussian fits to the disk
emission at all three PACS wavelengths.  This value of the inclination
is consistent with constraints from planetary modeling in the system
as well as with the inclination of the star itself, suggesting that
the system has strong alignment between all three components: star,
planets and disk. The agreement between the disk position angle and
the longitude of ascending nodes of the planets establishes that the
system is coplanar.

There is no evidence of asymmetric structure in the HR 8799 disk. The
symmetric models fit the disk so well that we estimate that, if
substructure is present, individual unresolved clumps could be no
brighter than 1\% of the total flux in the symmetric distribution of
emission.  Based on this limit on brightness asymmetry in the disk
emission, and comparable constraint from the similarity in the disk
radial separation of HR 8799 with Fomalhaut, we estimate that the disk eccentricity must be $< 0.1$.

\acknowledgements
We gratefully acknowledge the thorough report provided by our referee.  BCM, MB and HBF acknowledge the support of a Discovery Grant and a Discovery Accelerator Supplement from the Natural Science and Engineering Council (NSERC) of Canada.
MCW and GK are grateful for support from the European Union through ERC grant number 279973.  A portion of this work was performed under the auspices of the U.S. Department of Energy by Lawrence Livermore National Laboratory under Contract DE-AC52-07NA27344.

\end{document}